\documentclass{article}
\usepackage{xspace,fullpage,amsthm,amsmath,amssymb}
\usepackage{subfigure}
\usepackage[ansinew]{inputenc}
\usepackage{subfigure}

\usepackage[dvips]{epsfig}
\DeclareMathOperator{\polylog}{polylog}
 
\DeclareMathOperator{\discrep}{Disc}

\newcommand{\rounddown}[1]{\lfloor #1 \rfloor}
\newcommand{\roundup}[1]{\lceil #1 \rceil}

\newcommand{\minstop}{\ensuremath{\textsc{MinStop}}\xspace}
\newcommand{\mincstop}{\ensuremath{\textsc{MinCStop}}\xspace}

\newcommand{\minatrap}{\ensuremath{\textsc{MinTrap$_\mathrm{AWGN}$}}\xspace}
\newcommand{\minetrap}{\ensuremath{\textsc{MinTrap$_\mathrm{AWGN-elem}$}}\xspace}
\newcommand{\minzptrap}{\ensuremath{\textsc{MinTrap$_\mathrm{ZP}$}}\xspace}
\newcommand{\minagaltrap}{\ensuremath{\textsc{MinTrap$_\mathrm{GA}$}}\xspace}
\newcommand{\mld}{\ensuremath{\textsc{MaxLikeDecode}}\xspace}
\newcommand{\mtdm}{\ensuremath{\textsc{MaxThreeDimMatch}}\xspace}
\newcommand{\minctrap}{\ensuremath{\textsc{MinTrap$_{\mathrm{AWGN-maj}}$}}\xspace}
\newcommand{\mincw}{\ensuremath{\textsc{MinCodeword}}\xspace}
\newcommand{\minsc}{\ensuremath{\textsc{MinSetCov}}\xspace}
\newcommand{\minhs}{\ensuremath{\textsc{MinHitSet}}\xspace}
\newcommand{\minscinter}{\ensuremath{\textsc{MinSetCovInterOne}}\xspace}
\newcommand{\minvc}{\ensuremath{\textsc{MinVertCov}}\xspace}

\newcommand{\mingood}{\ensuremath{\textsc{MinGood}}\xspace}
\newcommand{\zigzag}{\ensuremath{\textup{\textsf{ZigZag}}}\xspace}
\newcommand{\orgate}{\ensuremath{\textup{\textsf{OrGate}}}\xspace}
\newcommand{\ballast}{\ensuremath{\textup{\textsf{Ballast}}}\xspace}

\newtheorem{proposition}{Proposition}
\newtheorem{theorem}{Theorem}
\newtheorem{corollary}[proposition]{Corollary}
\newtheorem{definition}{Definition}
\newtheorem{lemma}[proposition]{Lemma}

\title{On the Hardness of Approximating Stopping and Trapping Sets
\footnote{Part of the results were presented at the 2007
Information Theory Workshop, Lake Tahoe. The work was supported in
part by the NSF Grant CCF 0644427, the NSF Career Award, and the
DARPA Young Faculty Award of the second author.}}

\author{
Andrew McGregor\thanks{Microsoft Research, Silicon Valley Campus. Email: {\tt amcgreg@microsoft.com}.}
 \and
Olgica Milenkovic\thanks{Dept.\ of Electrical and Computer Engineering, University of Illinois, Urbana-Champaign. Email: {\tt milenkov@uiuc.edu}.}}
 \date{}

\begin{document}

\maketitle
\begin{abstract} We prove that approximating the size of stopping
and trapping sets in Tanner graphs of linear block codes, and more
restrictively, the class of low-density parity-check (LDPC) codes,
is NP-hard.
The ramifications of our
findings are that methods used for estimating the height of the
error-floor of moderate- and long-length LDPC codes based on
stopping and trapping set enumeration cannot provide accurate
worst-case performance predictions.
\end{abstract}

\section{Introduction}

In the past decade, the search for efficient and near-optimal
decoding algorithms for linear block codes culminated with the
rediscovery and generalization of the notion of sparse codes and
iterative message passing algorithms. Although Maximum Likelihood
(ML) decoding of linear block codes is NP-hard~\cite{berlekamp78},
iterative decoders can approach the Shannon limit of reliable
communication with polynomial time complexity, provided that they
operate on codes with long length that have sparse parity-check
matrices, also known as LDPC codes~\cite{gallager63}. Decoding is
achieved via message passing on  the \emph{Tanner graph} of the
code, a suitably chosen bipartite
graphical representation of the code which contains a very small
number of edges. On such graphs, probabilistic inference of the form of
iterative message passing is known to have linear complexity in
the code length.

The performance of linear block codes under iterative decoding,
and the performance of LDPC codes in particular, depends on the
structural properties of their chosen Tanner graphs. For each
channel-decoder pair, there exist vertex configurations in the
code graph on which the given iterative decoder fails. For some
frequently encountered Discrete Memoryless Channels (DMCs), such
configurations are known as \emph{near-codewords}~\cite{postol},
\emph{trapping and stopping sets}~\cite{di02,rich03},
\emph{pseudocodewords}~\cite{wiberg,koetter05}, and
\emph{instantons}~\cite{stepanov05}.

It is known that ML decoders fail when transmission errors are
confined to Tanner graph configurations containing codewords,
while iterative decoders usually fail to make correct decisions on
(strictly) larger sets of configurations. For example, iterative
edge-removal (ER) decoders for signalling over the Binary Erasure
Channel (BEC) fail on stopping sets~\cite{di02}, a subset of which
are the codewords themselves. For the Additive White Gaussian
Noise (AWGN) channel and sum-product decoding, failures arise due
to subsets of vertices in the code graph that have similar
structural properties as codewords, and are consequently termed
\emph{near-codewords}~\cite{postol}. As a result, iterative
decoders exhibit sub-optimal performance compared to ML decoders,
and this performance loss most frequently manifests itself in
terms of the emergence of \emph{error-floors} in the
Bit-Error-Rate (BER) curve of the code.

The error-floor phenomena is a problem of focal importance in the
theory of iterative decoding, since many practical applications of
codes on graphs require extremely low operational BERs. Since such
low BERs are well beyond the scope of current Monte-Carlo
simulation techniques, several methods were proposed for
estimating the height of the error-floor through enumerating small
stopping and small trapping sets~\cite{rosnes07}, and exploring
dominant instantons~\cite{rich03,stepanov05}. These techniques
operate fairly accurately for codes of very short and moderate
length and small minimum pseudoweight, but they are time
consuming, and no rigorous analytical study of the performance of
these search procedures is known.

Recently, it was shown that the problem of finding the smallest
stopping set in an arbitrary code graph is NP-hard to approximate
up to a constant term~\cite{KC06}. In~\cite{wang06}, it was shown
that finding the smallest \emph{$k$-out set}, which represents a
straightforward generalization of the notion of a stopping set, is
NP-hard as well. Despite the fact that $k$-out sets may lead to
decoding failures similar to those caused by trapping sets, the
results in~\cite{wang06} do not capture the fact that trapping
sets are usually characterized in terms of two parameters.
Furthermore, the notion of a trapping set is meaningful only in
conjunction with a fixed decoding method. Finally, no hardness
results for approximating $k$-out sets or more general trapping
sets are currently known.

The main contributions of our work are three-fold. First, we
improve upon the hardness results for approximating stopping sets,
presented in~\cite{KC06}. Furthermore, we introduce the notion of
a \emph{cover stopping set}, and show that the problem of finding
such a set of smallest cardinality in an arbitrary Tanner graph is
NP-hard. Second, we provide a set of new results regarding the
hardness of finding trapping sets for Gallager A decoder
(GA)~\cite{bazzietal04}, the Zyablov-Pinsker (ZP)
decoder~\cite{zyablov76,zigangirov07}, and the product-sum
decoder. The third, and most important finding presented in the
paper is that these hardness results carry over to the case of
LDPC code graphs (provided that the notion of ``low-density'' is
properly defined). We discuss the impact of these findings on the
accuracy of estimating the error-floor based on trapping set
enumeration techniques. In addition, we give a brief overview of
the theory of fixed parameter tractability (FPT), and show that
the minimum cover stopping set problem is FPT.

The paper is organized as follows. Section~\ref{sec:setup}
introduces the trapping set structures under investigation, as
well as their corresponding decoding algorithms.
Section~\ref{sec:hard} provides a brief overview of a class of
NP-hard problems that are used in the reduction proofs of our main
results. Section~\ref{sec:main} contains theorems regarding the
hardness of approximating classes of trapping sets, while
Section~\ref{sec:ldpc} specializes these results for the class of
sparse code graphs and short code lengths. In
Section~\ref{sec:errorfloor} we briefly comment on the accuracy of
error-floor estimation procedures relying on exhaustive trapping
set enumeration techniques. In Section~\ref{sec:fpt}, we describe the
notion of fixed parameter tractability and its implications for
stopping and trapping set size estimation. Concluding remarks are
given in Section~\ref{sec:conc}.

\begin{figure}
\begin{center}
{\centering \subfigure[Stopping Set.]
{ \includegraphics[scale=0.25,angle=-90]{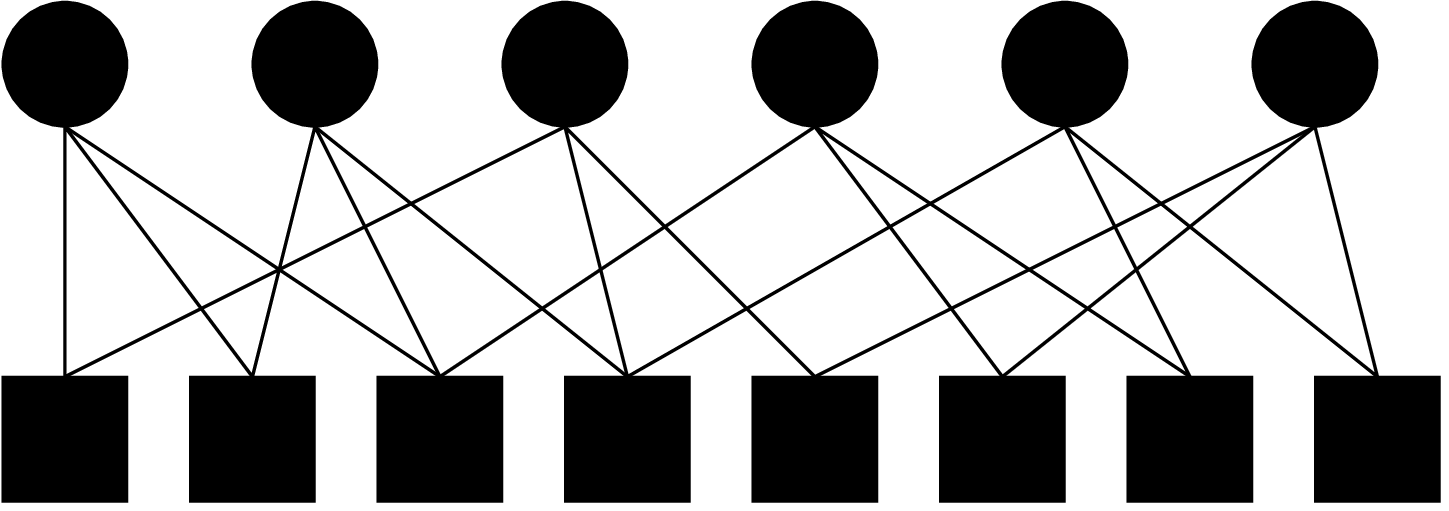} }}
\hspace{0.5in} {\centering \subfigure[ZP Trapping
Set.] {\includegraphics[scale=0.25,angle=-90]{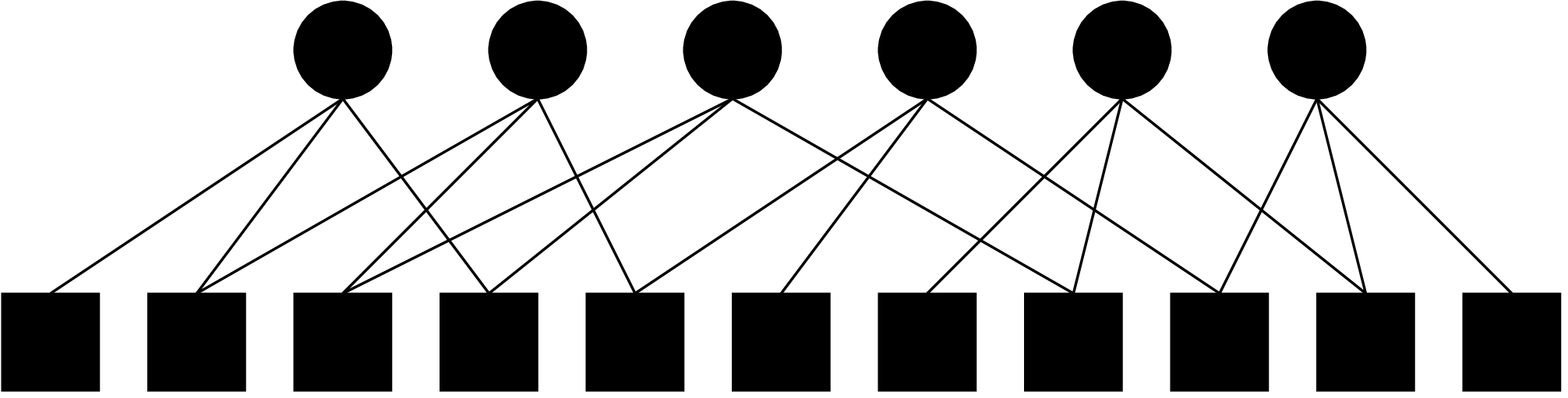}}}
\end{center}
\vspace{-8pt}
\caption{\label{case12}Examples of Stopping and ZP-Trapping Sets.}
\end{figure}

\begin{figure}
\begin{center}
{\centering \subfigure[AWGN $(a,b)$-Trapping Set.]
{\hspace{.2in}\includegraphics[scale=0.25,angle=-90]{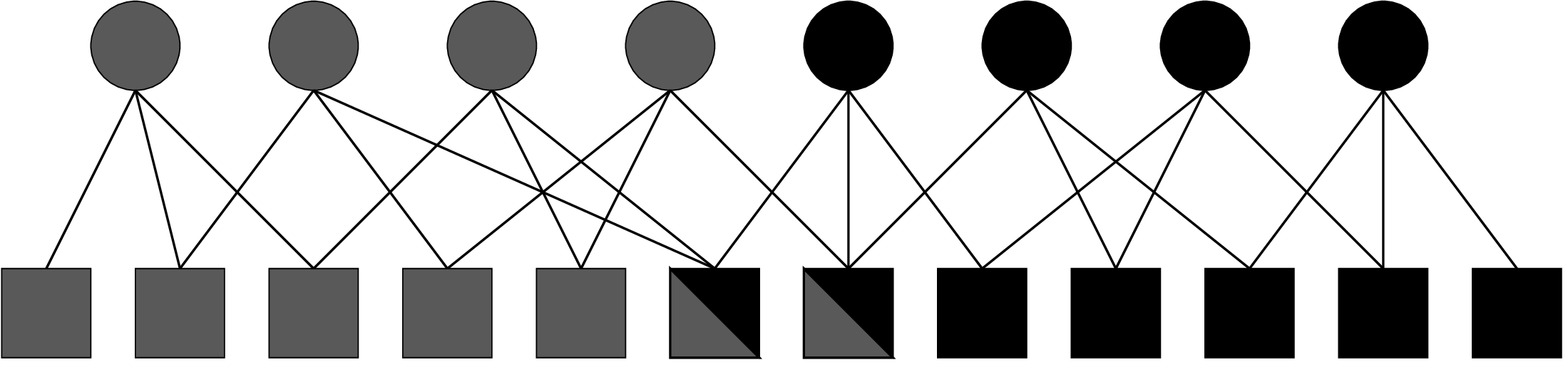}\hspace{.2in}}}
\hspace{0.1in} {\centering \subfigure[AWGN Elementary $(a,b)$-Trapping Set.] {\hspace{.2in} \includegraphics[scale=0.25,angle=-90]{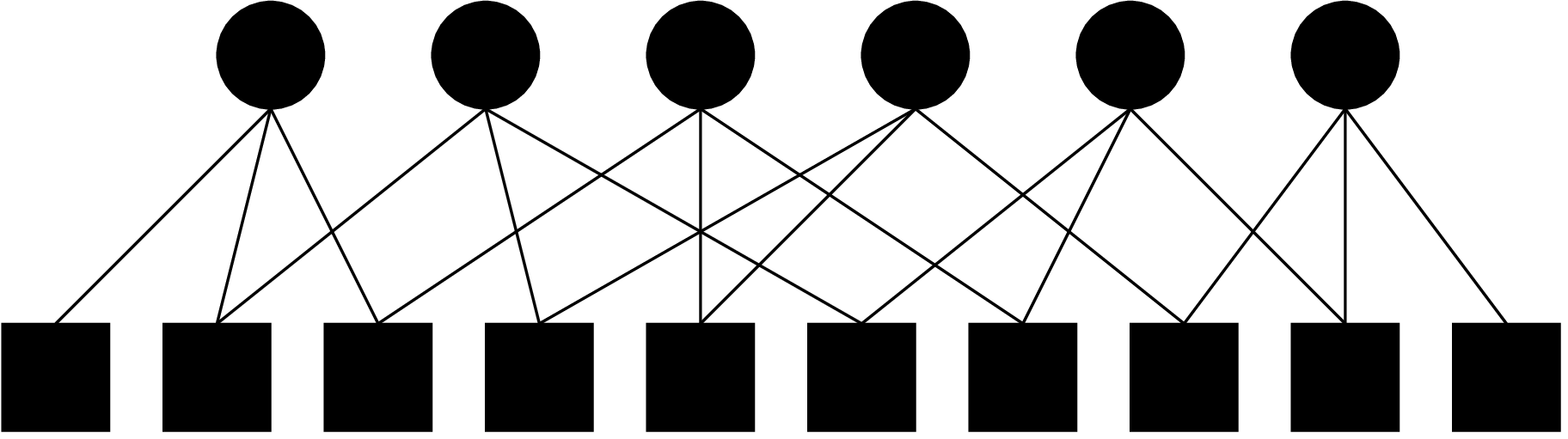} \hspace{.2in}}}
\end{center}
\caption{\label{case34} Examples of AWGN Trapping Sets.}
\end{figure}

\section{Definitions and Problem Formulation} \label{sec:setup}


A binary, linear $[n,k,d]$ code $\mathcal{C}$ is a $k$-dimensional
vector subspace of an $n$-dimensional vector space $F_2^n$. The
generator matrix $M$ of the code $\mathcal{C}$ is a $k \times n$
matrix of full row-rank, with rows that correspond to basis
vectors of the subspace. The parity-check matrix $H$ of
$\mathcal{C}$ is the generator matrix of the null-space of the
code. The matrix $H$ defines a bipartite graph $G=(L\cup R,E)$,
with columns of $H$ indexing the \emph{variable nodes} in $L$, and
the rows of $H$ indexing the \emph{check nodes} in $R$. For $i \in
L$ and $j \in R$, $(i,j) \in E$ if and only if $H_{i,j}=1$. 
The graph $G$ is called the Tanner
graph of $\mathcal{C}$ with parity-check matrix $H$. If the
parity-check matrix of a code contains only a ``small'' number of
non-zero entries, i.e.,  it is sparse, then the corresponding code
is called a Low-Density Parity-Check (LDPC) code. A precise
definition of the notion ``small'' will be given in
Section~\ref{sec:ldpc}.

For the remaining definitions in this section we need to introduce the following notation and definitions. For $S \subset L$, the notation $\Gamma(S)$ is reserved for the
set of neighbors of $S$ in $R$. $G_S$ denotes the induced subgraph for $S\subset L$ which is defined as the graph on nodes $S\cup \Gamma(S)$ with edges $\{(u,v):u\in S, v\in \Gamma(S)\}$. Equivalently, $G_S$ is the Tanner graph of the punctured parity-check
matrix of the code, consisting of the columns indexed by $S$.
For any graph $G'$, $V(G')$ denotes the set of nodes of $G'$ and $E(G')$ 

Iterative decoders are a class of inference algorithms that
operate on Tanner graphs of codes. These decoders are known to
compute the maximum likelihood estimates of variables only on
Tanner graphs free of cycles. Nevertheless, when applied to LDPC
codes that contain cycles, they can approach the Shannon limit on
optimal performance with complexity linear in the length of the
code.

The messages passed between vertices of the Tanner graphs during
iterative decoding depend on the characteristics of the
transmission channel, and there usually exist many different
iterative decoding methods that can be used for the same channel.
For various decoder architectures specialized for the BEC, BSC,
and AWGN channel, the interested reader is referred
to~\cite{richardson01}. For clarity of the future exposition, we
briefly describe three of these procedures: the edge-removal (ER)
algorithm, the Zyablov-Pinsker (ZP) bit-flipping
method~\cite{di02,zyablov76,zigangirov07}, and the regular
Gallager A algorithm~\cite{vasic07}. The first algorithm operates
on outputs of the BEC, while the second two are designed for the
BSC. A detailed description of different decoding procedures for
signalling over the AWGN channel can be found in~\cite{rich03}.

The ER algorithm is used for codes transmitted over the BEC
channel, where the input to the channel is a vector $c_1 c_2
\ldots c_n \in \mathcal{C}$, and the output is a vector $v_1
v_2\ldots v_n$ over the symbol alphabet $\{{0,1,e\}}$. For a BEC
channel with erasure probability $p$, one has $\Pr[v_i=c_i]=1-p$,
and $\Pr[v_i=e]=p$. The ER algorithm assigns to each vertex $i$ in
$L$ of the Tanner graph of $\mathcal{C}$ the symbol $v_i$. It then
iteratively searches for vertices in $R$ adjacent only to one $e$
symbol in $L$. Due to the even-parity restriction, the
corresponding $c_i$ value for such a symbol can be uniquely
determined. The decoder terminates either when the correct
codeword is recovered or if every every parity-check vertex
connected to one $e$ symbol is connected to at least two such
symbols. In the latter case, we say that the decoder failed on a
\emph{stopping set}.

\begin{definition}[BEC Stopping Sets]
Given a bipartite graph $G=(L\cup R,E)$, we say that $S\subset L$
is a stopping-set if the degree of each vertex in $\Gamma(S)$ in
the induced subgraph $G_S$ is at least two.
\end{definition}

Of independent interest is the problem of determining the size of
the smallest stopping set $S$ such that $\Gamma(S)=R$, i.e., the
smallest set of vertices that covers \emph{each} check node in $R$
at least twice. We refer to such a set as the \emph{cover}
stopping set. If symbols corresponding to a cover stopping set are
erased, then the decoding process terminates before proceeding
with the first iteration, and no erasure can be corrected.

Assume next that the Tanner graph of $\mathcal{C}$ is
left-regular, with degree $\ell$. For a BSC channel with error
probability $p$, the word $v_1 v_2 \ldots v_n \in \{{0,1\}}^n$ and
$\Pr[v_i=c_i]=1-p$, and $\Pr[v_i=\bar{c}_i]=p$. In the first
iteration of ZP-decoding, the decoder scans for received symbols
$v_i$ that are connected to $\ell$ unsatisfied parity-check
equations. If symbols with such a property are encountered, the
decoder flips their values sequentially. The procedure is repeated
for vertices with $\ell-1$, $\ell-2$, ...,
$\ell-\lfloor(\ell-1)/2\rfloor$ unsatisfied check-equations. The
decoder terminates by either recovering the correct codeword or by
encountering a word for which each symbol is included in less than
$\ell-\lfloor(\ell-1)/2\rfloor$ unsatisfied check-equations. In
the latter case, we say that the decoder failed on a \emph{ZP
trapping set}.

\begin{definition}[BSC ZP-Trapping Sets]
Let $G=(L\cup R,E)$ be a left-regular bipartite graph with degree
$\ell$. We say that $S\subset L$ is a \emph{ZP-trapping set} if the
induced subgraph $G_S$ is such that all vertices in $S$ are
connected to less than $\ell-\lfloor(\ell-1)/2\rfloor$ odd degree
vertices in $G_S$.
\end{definition}

\begin{figure}[tp]
\centering \epsfig{file=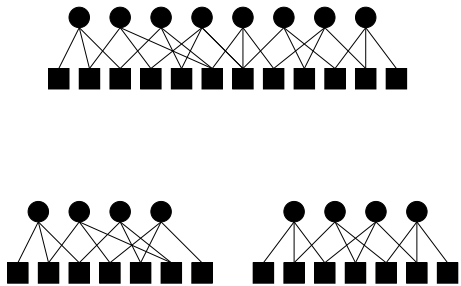, angle=-90, width=6.0cm}
\caption{Example of Majority Trapping Set.} \label{case5}
\end{figure}

Another frequently used iterative decoding algorithm for signaling
over the BSC that has a complete characterization of trapping sets
is the Gallager A algorithm for regular codes with left vertex
degree $\ell=3$. The decoding rule is straightforward: unless all
incoming massages to a variable node are identical, the variable
node transmits its received symbol. Otherwise, the node transmits
the consensus vote. On the other hand, the check-nodes pass on
their parity estimates to their neighboring variable nodes.

\begin{definition} [BSC GA-Trapping Sets]
Let $G=(L\cup R,E)$ be a bipartite graph with left-degree three,
such that all vertices in $R$ have degree $r>3$. Let $T\subset L$ and let $G_T$ be the subgraph of $G$ induced by
$T$. Let $O=\{v\in \Gamma(T): \deg_{G_T}(v)\mbox {odd}\}$. We say
$T$ is a \emph{GA-trapping set} with parameter $a$ if $|O|=a$ and if $|\Gamma(u)\cap O|\leq 1$ for each $u\in T$ and no two
checks in $O$ have a common neighbor in $L \setminus T$.
\end{definition}

For the AWGN channel, and message-passing algorithms, no precise
analytic characterization of failing configurations is known.
Extensive computer simulations~\cite{rich03,ml} show that errors
are usually confined to \emph{near codewords}, also known as
trapping sets or instantons. Roughly speaking, trapping sets
resemble codewords in so far that they result in a very small
number of unsatisfied check equations (for codewords, this number
equals zero). We focus our attention on three such configurations,
defined below.

\begin{definition}[AWGN $(a,b)$-Trapping Sets]
Given a bipartite graph $G=(L\cup R,E)$, we say that $S\subset L$
is an \emph{$(a,b)$-trapping set} if $|S|=a$ and the induced subgraph
is such that $\Gamma(S)$ has exactly $b$ vertices of odd degree.
Similarly, we say that $S\subset L$ is an \emph{elementary $(a,b)$-trapping set} if $b$ vertices in $\Gamma(S)$ have degree one,
and $|\Gamma(S)|-b$ vertices have degree two.
\end{definition}

\begin{definition}[AWGN Majority Trapping Set]
Given a bipartite graph $G=(L\cup R,E)$ we say $S\subset L$ is
\emph{good} if the induced subgraph $G_S$ is such that the majority of
vertices of $G_S$ in $\Gamma(S)$ have even degree. $T$ is a \emph{majority trapping set} if $T$ and $L\setminus
T$ are both good.
\end{definition}

Examples of Tanner graphs including stopping sets, ZP-trapping
sets, as well as AWGN trapping sets are shown in
Figures~\ref{case12},~\ref{case34}, and~\ref{case5}, respectively.
Circles denote variable nodes in $L$, while squares denote check
nodes in $R$ of the Tanner graph $G(L\cup R,E)$.

\paragraph{Complexity Theory:} A problem belongs to the class NP if it can be solved in
polynomial time by a non-deterministic Turing machine.
Alternatively, the complexity category of decision problems for
which answers can be checked for correctness using a certificate
and an algorithm with polynomial running time in the size of the
input is known as the NP class.
A problem is NP-hard if the existence of a deterministic
polynomial time algorithm for the problem would imply the
existence of deterministic polynomial time algorithms for every
problem in NP. This consequence is widely believed to be false,
and hence determining that a problem is NP-hard is a very strong
indicator that the problem in computational intractable, i.e., no
deterministic, polynomial time algorithm exists for the problem.


For optimization problems, there exists a large body of work that
considers approximate solutions rather than exact
solutions~\cite{Vazirani}. When minimizing a function subject to
constraints, we say an algorithm is an $\alpha$-approximation
algorithm if it always returns a solution whose value is at most a
factor $\alpha$ greater than the value for the optimal solution.
For some NP-hard problems, it is possible to show that it is also
NP-hard to $\alpha$-approximate the problem.
For a more thorough treatment of these and other subjects in
complexity theory, the interested reader is referred
to~\cite{garey79}.

\paragraph{Our Results:} We are concerned with the worst-case computational complexity of
the following problems.

\begin{enumerate}
\item $\minstop$: Find a stopping set of minimum cardinality.
\item $\mincstop$: Find a cover stopping
set of minimum cardinality . 
\item $\minzptrap$: Find a ZP-trapping set of minimum
cardinality. 
\item $\minagaltrap$: Given $a$, find a GA-trapping set of minimum cardinality. 
\item $\minatrap$: Given $a$, find an $(a,b)$-trapping set with minimum parameter $b$.
\item $\minetrap$: Given $b$, find an $(a,b)$-elementary trapping set with minimum parameter $a$.
\item $\minctrap$: Find a majority trapping set of minimum cardinality.
\end{enumerate}

We show that there are no polynomial time algorithms for any of the above problems under standard complexity assumptions. Furthermore,  there are no polynomial time algorithms that even approximate the optimal solutions to a guaranteed precision. 
Many of these hardness results  also apply when we restrict our attention to  Tanner graphs that correspond to  LDPC codes. Our proofs
can all be cast as reductions from the NP-hard \emph{Minimum Set
Cover}, \emph{Minimum
Distance}~\cite{vardy-1-97,vardy-2-97,DMS03}, and \emph{Maximum
Three-Dimensional Matching} problems~\cite{berlekamp78}. These,
and some other relevant problems subsequently referred to, are
briefly described in the following section.

\section{A Class of NP-hard problems}
\label{sec:hard}

For completeness, we provide known NP hardness and approximation
results for a class of combinatorial optimization problems that
will be used in the proofs of
Sections~\ref{sec:hard},~\ref{sec:main}, and~\ref{sec:ldpc}. Most
of the results presented in this section are available
at~\cite{viggo}.

\begin{enumerate}
\item \textbf{The Minimum Set Cover Problem, \minsc:} Given a set
of sets ${\mathcal S}=\{S_1, \ldots, S_a\}$ of $[b]$, find
${\mathcal S}'\subset {\mathcal S}$ of minimum cardinality such
that $\cup_{S\in {\mathcal S}'} S= [b]$. It is NP-hard to $c\log
N$-approximate \minsc \cite{RazS97} for some $c$ where $N$ is the description length of the problem.
Even  in the case that $|S_i \cap S_j| \leq 1$, for $1 \leq i<j
\leq a$, it can be shown that there exists no polynomial time $c
\log N$-approximation algorithm unless $NP \subset ZTIME(N^{O(\log
\log N)})$~\cite{kumar} where $ZTIME(t)$ denotes the class of
problems that have a probabilistic algorithm with expected running
time $t$ and with zero error probability.

%

\item \textbf{The Minimum Hitting Set Problem, \minhs:} Given a
set of subsets ${\mathcal S}=\{S_1, \ldots, S_b\}$ of $[a]$, find
a set $S'$ of smallest cardinality, such that $|S' \cap S_i| \geq
1$, for all $i=1,2,\ldots, b$. The \minhs problem is equivalent to
the \minsc problem~\cite{ausiello80} and as a consequence it is
also NP-hard to $(c\log N)$-approximate \minhs \cite{RazS97} for
some $c>0$. In the case when $|S_i|=2$ for all $i\in [b]$ the
problem is often called the vertex cover problem \minvc. The vertex cover
problem, even when we have $|\{i:j\in S_i\}|\leq 3$ is NP-hard to
approximate up to some constant $\alpha>1$.

%
%
%

\item \textbf{The Maximum Three-Dimensional Matching Problem,
\mtdm:} Given a set $T \subset X \times X \times X$, determine if
a set $S \subset T$ of size $|X|$ exists such that no elements in
$S$ agree in any coordinate. This decision problem is NP-hard even
if no element of $X$ appears more than $3$ times in the same
coordinate of sets from $T$ \cite{garey79}.

\item \textbf{The Maximum Likelihood Decoding Problem, \mld:}
Given a code $\mathcal C$ specified by an $m\times n$ parity-check
matrix $H$ (we may assume $H$ has linearly independent rows), a
vector $s \in F_2^m$, and an integer $\omega>0$, determine if
there is a vector $x \in F_2^n$ with weight bounded from above by
$\omega$ and such that $H\,x^T=s$. The \mld problem is NP-hard to
approximate within any constant factor~\cite{arora97}.

\item \textbf{The Minimum Weight Codeword Problem, \mincw:} Given
a code $\mathcal C$ specified by an $n\times k$ generator matrix
$M$ of full row-rank, find the smallest weight of a non-zero
codeword. The \mincw problem is not approximable within any
constant factor unless $NP\subset RP$, where RP is the set of
decision problems for which there exists a randomized algorithm
that is always correct on no instances and correct with
probability 1/2 on yes instances.
\end{enumerate}

\section{Hardness of Approximation Results} \label{sec:main}


\subsection{Hardness of Approximation for \minstop}

We start by showing that  \minstop is not approximable within
$o(\log\,N)$, where $N$ denotes the description length of the
problem, unless $P=NP$. This results improves upon the finding
in~\cite{KC06}, where the weaker claim that \minstop cannot be
approximated within any positive constant was proved. This
improvement is a consequence of the fact that our proof relies on
reduction from the \minsc, rather than the \minvc
problem~\cite{KC06}.

\begin{theorem} \label{th:stop}
There exists a constant $c>0$ such that it is NP-hard to $(c\log
N)$-approximate \minstop.
\end{theorem}
\begin{proof}
The proof is by a reduction from \minsc.  Let $b=\left |
\cup_{i\in [a]} S_i\right |$, and without loss of generality,
assume that $S\subset [b]$, for each $S\in {\mathcal S}.$ Form a
bipartite graph $G=(L\cup R,E)$ with $L= \{u_1, ... u_a, x, y\}$,
$ R= \{v_1, ... v_b, w_1, ..., w_a, z\}$, and edges \[E=
\{(u_i,v_j): j\in S_i\}\cup \{(u_i,w_i): i\in [a]\}\cup
\{(x,v):v\in R\}\cup \{(y,v):v\in \{w_1, ..., w_a, z\}\} \enspace
.\]
An illustration of this graphical structure is given in Figure 4.

We show that $G$ has stopping distance $2+t$ if and only if the
minimum set cover is of size $t$. Since there is no polynomial
algorithm returning an $c\log N$ approximation for \minsc unless
$P=NP$ (for some sufficiently small $c>0$), this establishes the
theorem.

Let $S$ be a stopping set. Consequently,
\begin{enumerate}
\item If $ \left (x\in S \mbox{ or } y\in S \right )$, then $\left
( x\in S \mbox{ and } y\in S \right )$ since otherwise
$d_{G_S}(z)=1$. \item If $x\in S$ then $u_i\in S$ for some $i$
since otherwise $d_{G_S}(v_j)=1$ for some $j\in [b]$. \item If
$u_i\in S$ then $\left ( x\in S  \mbox{ or }  y\in S \right )$
since otherwise $d_{G_S}(w_i)=1$.
\end{enumerate}

Therefore, if  $S$ is non-empty $x, y, u_i\in S$ for some $i\in
[a]$. But then $d_{G_S}(v_j)\geq 2$ for $j\in S_i$. However this
means that for all $j\in [b], \ d_{G_{S}\setminus
\{x,y\}}(v_j)\geq 1$. Therefore, $S$ being a stopping set implies
that  the included $u_i$ nodes correspond to a covering of $[b]$.
The nodes corresponding to a covering of $[b]$, in addition to $x$
and $y$, form a stopping set, since every node on the right hand
side ($R$) is in the neighborhood and has degree at least two.
Hence the size of the minimum stopping set of $G$ is exactly 2 plus the size of the minimum set cover.
\end{proof}

\begin{figure}[tb]
\centering \epsfig{file=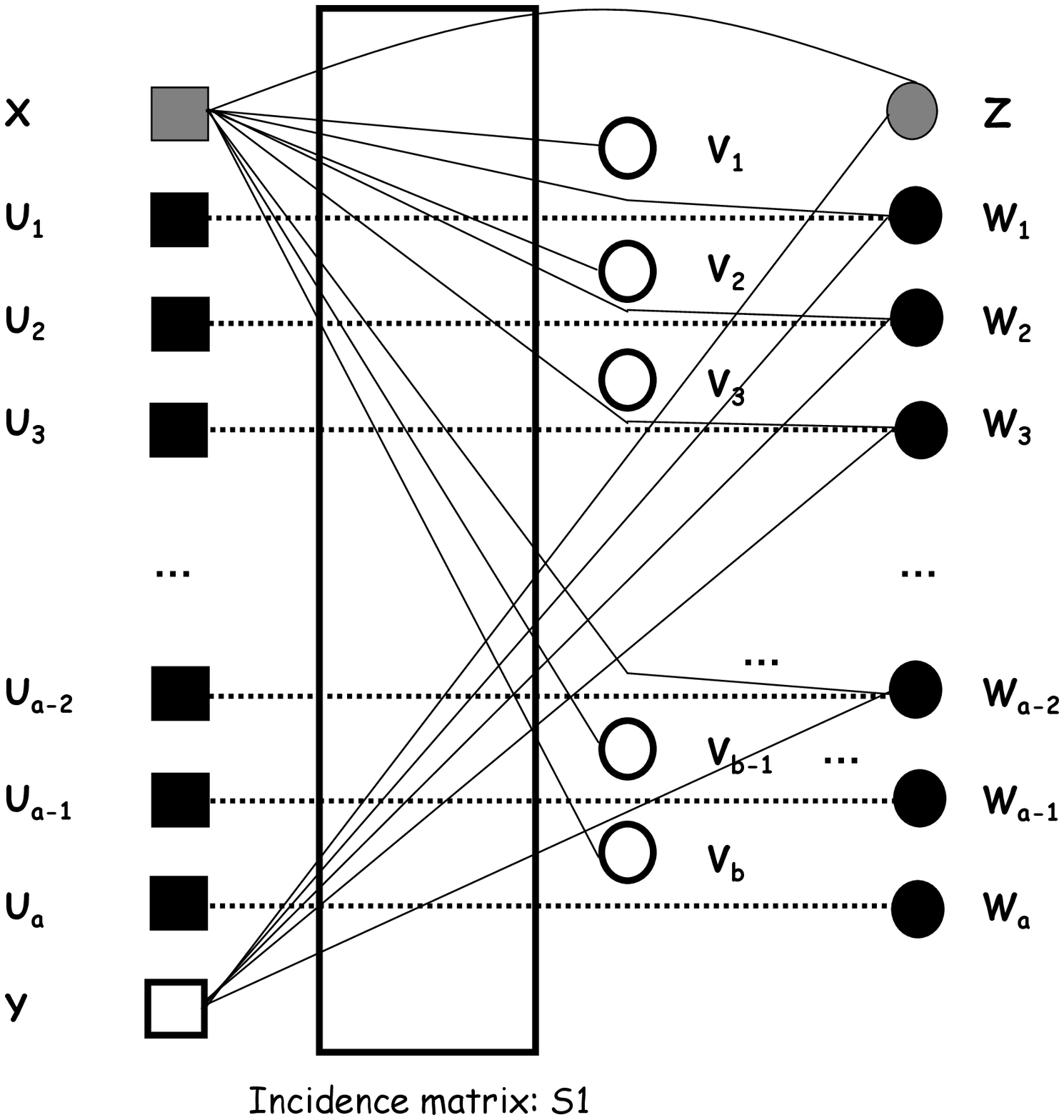, angle=-90,
width=5.5cm} \caption{Reduction from \minsc to \minstop.}
\label{fig:stopred}
\end{figure}



\paragraph{\mincstop:}
The proof of Theorem~\ref{th:stop} also implies that there exists
a $c>0$ such that it is NP-hard to $(c\log n)$-approximate
\mincstop. This is a consequence of the fact that the family of
hard instances considered all had the property that the
neighborhood of all stopping sets was all the check nodes. We next
show that there exists a deterministic, polynomial-time, $O(\log
n)$-approximation algorithm for \mincstop. This follows because we
can relate \mincstop to \minhs as follows.

For each $r \in R$, create a set of sets $S_r$ that consists of
all $(|\Gamma(r)|-1)$-subsets of $\Gamma(r)$. For example, if
$\Gamma(r)=\{{a,b,c,d\}}$, then $
S_r=\{{(a,b,c),(a,b,d),(a,c,d),(b,c,d)\}}. $ Let ${\mathcal
S}=\{S_r:r\in R\}$. Then $Q \subset L$ is a hitting set for
$\mathcal S$ iff it is a cover stopping set of $L$. This claim can
be proved in a straightforward manner: if $\mathcal S$ contains at
least one element, say $a$, from $\Gamma(r)$, then it must contain
at least two elements from the same set since otherwise, the
$(|\Gamma(r)|-1)$ set that does not contain $a$ will not be hit.

Consequently, any $\alpha$-approximation algorithm for \minhs can
also be used to obtain an $\alpha$-approximation algorithm for
\mincstop. For example the following simple greedy algorithm can
be shown to be an $O(\log n)$-approximation algorithm for
$\minhs$: At each step add the element that appears in the most
sets from $\mathcal S$ can remove these sets from $\mathcal S$ can
repeat until all the elements chosen appear in every set from
$\mathcal S$.
%
%
%

The greedy algorithm searches for cover stopping sets by going
through the list of variable nodes in decreasing order of their
degree, and it is straightforward to see that the algorithm
terminates after at most $(n-k)\, \delta_{\max}$ steps, where
$\delta_{\max}$ denotes the largest degree of any check node in
the Tanner graph of the code. As a consequence, this algorithm is
especially well suited for LDPC codes, to be formally defined in
Section~\ref{sec:ldpc}.

\paragraph{Hardness under Stronger Assumptions:} Under the assumption that $NP\not \subset
DTIME(N^{\polylog N})$, it was shown in~\cite{KC06} that there
exists no polynomial time approximation algorithm for \minstop
within $2^{(\log N)^{1-\epsilon}},$ for any $\epsilon>0$.

\subsection{Hardness of Approximation for \minzptrap, \minagaltrap, and \minatrap}
We show next that the problems \minzptrap, \minagaltrap, and \minatrap are
computationally at least as hard as the \mincw problem.

\begin{theorem} \label{minzp}
For any constant $\alpha$, there is no polynomial-time
$\alpha$-approximation algorithm for $\minzptrap$, unless $RP=NP$.
\end{theorem}
\begin{proof}
Recall that unless $RP=NP$, there is no polynomial time  \mincw
problem is $O(1)$-hard to approximate even under the restriction
that the Tanner graph of the code is left regular. This follows
directly from the results in~\cite{DMS03}.

Given a Tanner graph $G=(L\cup R,E)$ that is left regular say with
degree $\rounddown{(\ell-1)/2}+1$, for each node $u \in L$ create
$\ell-\rounddown{(\ell-1)/2}-1$ new nodes in $R$ each connected to
$u$. Call the new Tanner graph $G'$. Then any $S\subset L$ is a
ZP-trapping set in $G'$ iff $S$ is the support of a codeword in
$G$. Hence any $\alpha$-approximation algorithm for \minzptrap yields an $\alpha$-approximation algorithm for \mincw and the result follows.
\end{proof}

A very similar argument can be used to prove the following claim.

\begin{theorem} \label{minagal}
For any constant $\alpha$, there is no polynomial-time,
$\alpha$-approximation algorithm for $\minagaltrap$, unless
$RP=NP$.
\end{theorem}
\begin{proof}
Similarly as in the proof of Theorem~\ref{minzp}, create for each
node $u \in L$ one new node in $R$ each connected only to $u$.
Call the new Tanner graph $G'$. Then any $S\subset L$ is a
GA-trapping set in $G'$ iff $S$ is the support of a codeword in
$G$. This follows due to the fact that the first condition in the
definition of GA-trapping sets is identical to the ZP-restriction,
with $\ell=3$. The second condition in the definition of an
GA-trapping set is enforced automatically, since vertices in $L
\setminus S$ cannot be connected to odd-degree check nodes in
$G_{S}$ due to the fact that all such checks have degree one. Hence any $\alpha$-approximation algorithm for \minzptrap yields an $\alpha$-approximation algorithm for \mincw and the result follows.

\end{proof}

\begin{theorem} \label{mina}
For any constant $\alpha$, there is no polynomial-time,
$\alpha$-approximation algorithm for $\minatrap$, unless $RP=NP$.
\end{theorem}
\begin{proof}
The proof is by a reduction from \mincw, and follows along similar
lines as the proof of the above theorems. To this end, we
construct the Tanner graph $(L\cup R, E)$ of the dual code
$C^\perp$ where
$L= \{u_1, ... u_k\}, R= \{v_1, ... v_n\},$ and $E= \{(u_i,v_j):
M_{i,j}=1\}$ where $M$ denotes a generator matrix of the code of
full row-rank.
Note that for each $S\subset L$, $\Gamma(S)$ corresponds to a
codeword. Hence, if we have an $\alpha$-approx to the min-trapping
set problem for any $a$, then this gives an $\alpha$ approximation
algorithm to the minimum weight codeword problem by running
through all values of $a$ and taking the minimum of the resulting
$b$'s. But, since it is impossible to $O(1)$-approximate \mincw in
polynomial time unless $RP=NP$  \cite{DMS03}, it is impossible to
$O(1)$-approximate \minatrap in polynomial time unless $RP=NP$.
\end{proof}

\subsection{Hardness of Approximation for \minetrap}

\begin{theorem} \label{th:mine}
For any $\alpha$, it is NP-hard to $\alpha$-approximate
$\minetrap$.
\end{theorem}
\begin{proof}
The proof is based on showing that a polynomial time algorithm for
solving \minetrap can be used for solving the \mtdm problem, and
is based on similar arguments as those used for showing that \mld
is NP-complete~\cite{berlekamp78}. To this end, let us construct
the \emph{matching incidence matrix} $D$ as follows. Let the
collection of ordered triples be $T \subset X \times X \times X$,
where $|T|=t$, and $|X|=n$. Then $D$ is a $3\,n \times t$
dimensional zero-one matrix, with entries
\begin{equation}
\begin{aligned}
1 \leq i \leq n:\;\, D_{i,j}=1, \; \text{iff} \;\, x_j=i;\\
n+1 \leq i \leq 2n:\;\, D_{i,j}=1, \; \text{iff} \;\, y_j=i;\\
2n+1 \leq i \leq 3n:\;\, D_{i,j}=1, \; \text{iff} \;\, z_j=i.
\notag
\end{aligned}
\end{equation}
As an example, the matrix $D$ for the set of triples
\begin{equation}
\{{(1,2,2),(3,2,1),(2,3,1),(1,2,3),(2,3,3),(3,1,3)\}} \notag
\end{equation}
over $X=\{{1,2,3\}}$ has the form
\begin{equation}
D=\left(%
\begin{array}{cccccc}
  1 & 0 & 0 & 1 & 0 & 0 \\
  0 & 0 & 1 & 0 & 1 & 0 \\
  0 & 1 & 0 & 0 & 0 & 1 \\
  0 & 0 & 0 & 0 & 0 & 1 \\
  1 & 1 & 0 & 1 & 0 & 0 \\
  0 & 0 & 1 & 0 & 1 & 0 \\
  0 & 1 & 1 & 0 & 0 & 0 \\
  1 & 0 & 0 & 0 & 0 & 0 \\
  0 & 0 & 0 & 1 & 1 & 1 \\
\end{array}
\right). \notag
\end{equation}
The set of triples $\{{(1,2,2),(2,3,1),(3,1,3)\}}$ is a maximum
three-dimensional matching over the set $\{{1,2,3\}}$. Observe
that all rows in the sub-matrix of $D$ induced by the three columns
corresponding to these triples have Hamming weight one. This is a
consequence of the defining constraint of the \mtdm problem that
asserts that every element in $X$ appears at a given position of
the matching exactly once.

Assume next that there exists a polynomial-time,
$\alpha$-approximation  algorithm for the \minetrap problem.
Construct $D$ for a given matching problem, set $b=3\times n$, and
run the \minetrap algorithm on $D$. If the algorithm the algorithm
finds an elementary trapping set then it must have size $n$.
Consider the corresponding set of $n$ columns indexed by a set of
$n$ triples from $T$. Each row in the sub-matrix induced by the
triples has weight one, which follows from the definition of an
elementary trapping set. Consequently, these triples represent a
matching for $T$. This implies that no polynomial time algorithm
for the \minetrap problem exists, unless P=NP.
\end{proof}

\subsection{Hardness of Approximation for \minctrap}

First we prove a hardness of approximation result for  the problem
of finding the good set of minimum cardinality. Recall that a set
$S\subset L$ us good if the majority of nodes in $\Gamma(S)$ have even degree in $G_S$.
We call this problem \mingood. We will then use this to show a
hardness of approximation result for \minctrap.

Our proof uses a reduction from \mincw. Let $H$ be the $n\times
(n-k)$ parity check of some code. We may assume that the code
specified by $H$ includes at least one codeword in addition to the
zero vector. This gives rise to the graph $G'=(L'\cup R',E')$
where $L'=\{x_1, \ldots, x_n\}, R'=\{y_1,\ldots, y_m\}, $ and $
E'=\{(x_i,y_j):H_{i,j}=1\}.$
We will create a bipartite graph $G=(L\cup R,E)$ by augmenting
$G'$ with graphical objects termed ``\zigzag{}"s and
``\orgate{}"s. These graphical objects will ensure that the
minimum cardinality of a good set is approximately proportional to
the minimum weight of any codeword.

\subsubsection{The \zigzag}
For each $x\in L'$ we add a $\zigzag(x)$ structure. This structure
consists of $3(m-1)$ nodes, given by $ L(\zigzag(x))= \{v_1,
\ldots, v_{m-1}\}, R(\zigzag(x))= \{u_1, \ldots, u_{m-1}, w_1,
\ldots, w_{m-1}\},$ and edges,
\[E(\zigzag(x))= \{(u_i,v_i),
(v_i,w_i):i\in [m-1]\} \cup \{(v_i,w_{i+1}: i\in
[m-2]\}\cup\{(x,w_1)\} \ \] The intuition behind the $\zigzag(x)$
structure is that if $x$ is in the trapping set then the nodes
$L(\zigzag(x))$ will also be in the trapping set.
For a subgraph $G''$ of $G$, and $S\in L$ we define
\begin{eqnarray*}
 \discrep_S(G'')&=&|\{v\in \Gamma(S)\cap V(G''): d_{G_S}(v) \mbox{ even}\}|
 -|\{v\in \Gamma(S)\cap V(G''): d_{G_S}(v) \mbox{ odd}\}|.
 \end{eqnarray*}
%

\begin{lemma}\label{lem:zigzag}
For all $x\in S$, $\discrep_S(\zigzag(x))\leq 0$ and
$\discrep_S(\zigzag(x))=0$ iff $\zigzag(x)\cap L \subset S$.
\end{lemma}
\begin{proof}
Note that \[|\{v\in \Gamma(S)\cap V(\zigzag(x)): d_{G_S}(v) \mbox{
odd}\}|\geq |\{v\in S\cap V(\zigzag(x))| \] with equality iff
$L'\cap V(\zigzag(x))\subset S$ because each $v_i\in S$ is
connected to $w_i$ which has degree 1. But for any $S$,
\[|\{v\in \Gamma(S)\cap V(\zigzag(x)): d_{G_S}(v) \mbox{ even}\}|\leq |\{v\in S\cap V(\zigzag(x))| \ \]
with equality iff $L'\cap V(\zigzag(x))\subset S$.
\end{proof}

\subsubsection{The \orgate}

For each $y\in R'$ we add $\orgate(y)$, and let $\Gamma(y)\cap
L'=\{u_1, \ldots, u_{k'}\}$. Let $k=2^{\roundup{\log_2 k'}}$. The
construction $\orgate(y)$ consists two node sets $L(\orgate(y))$
and $R(\orgate(y))$. Consider a binary tree on the nodes $\{u_1,
\ldots, u_{k}\}$ where $u_{k'+i}=u_{k'}$ for $i\in [k-k']$. Then
$L(\orgate(y))$ consists of  nodes corresponding to the internal
nodes of the tree, i.e.~
\[L(\orgate(y))=\{v_{u_1\vee u_2}, \ldots, v_{u_{k-1}\vee u_k},
 v_{u_1\vee u_2\vee u_3\vee u_4}, \ldots,  v_{u_{k-3}\vee u_{k-2}\vee u_{k-1}\vee u_k},
 \ldots, v_{u_1\vee u_2\vee \ldots \vee u_k}
\}
\]
For each internal node $v$ with children $u$ and $w$, we add four
new check nodes $C(v):=\{c_1(v), c_2(v), c_3(v), c_4(v)\}$: all
are connected $v$, the first and third are connected to $u$ and
the first and second are connected to $w$. If $v$ is the root of
the tree, we also add one more new check node which is connected
only to $v$. We call this node $z$. Let $R(\orgate(y))$ be the set
of such nodes, and let $E(\orgate(y))$ be the set of such edges.
Finally, let
\[f(S,y)=\{v_{u_i\vee \ldots \vee u_j}\in L(\orgate(y)): |S\cap \{u_i, \ldots, u_j\}|\geq 1\}\enspace .\]

\begin{figure}[tb]
\begin{center}
\subfigure[$\orgate(y)$]{\qquad\scalebox{0.5}{\epsfig{file=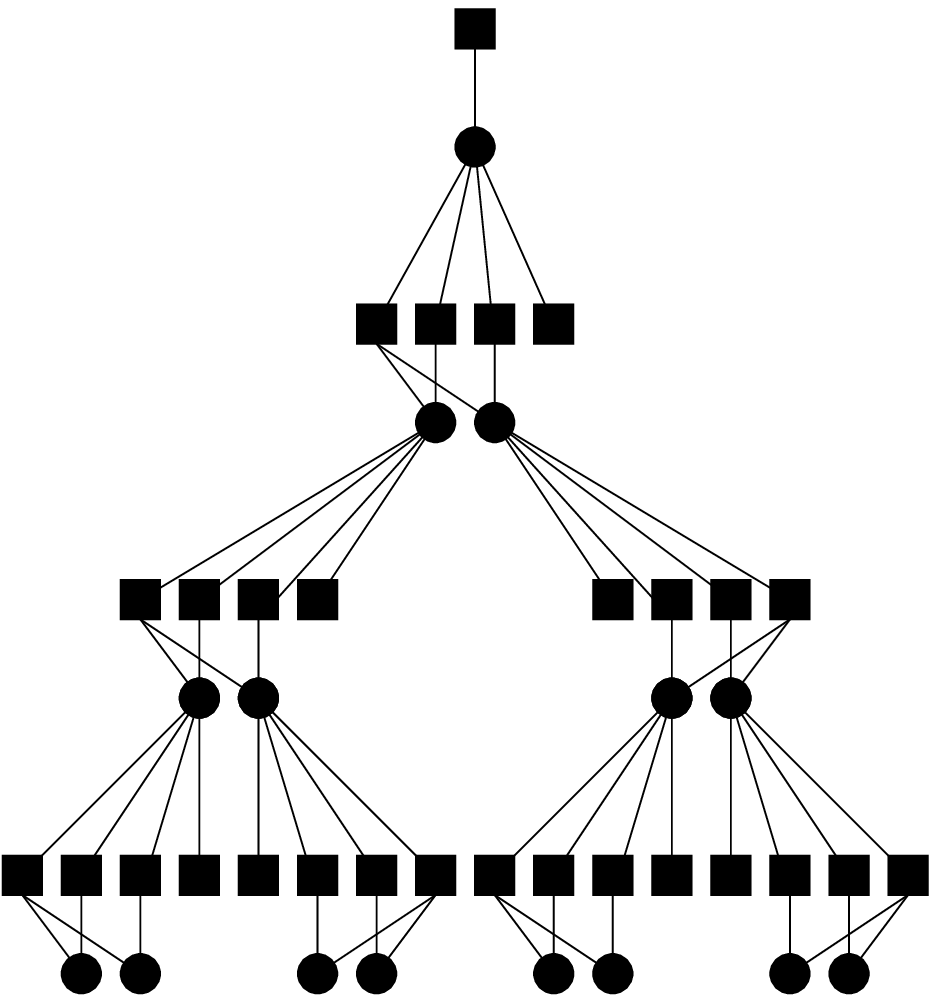}}
\qquad}
\subfigure[$\zigzag(x)$]{\qquad\scalebox{0.5}{\epsfig{file=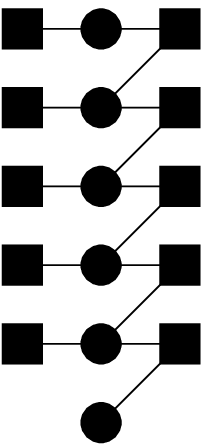}}
\qquad} \quad
\subfigure[$\ballast$]{\qquad\scalebox{0.5}{\epsfig{file=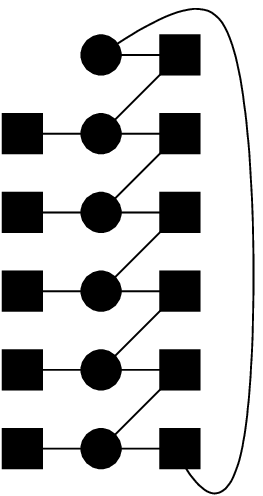}}
\qquad}
\caption{Reduction from \mincw to \minctrap.} \label{fig:orgate}
\end{center}\end{figure}


\begin{lemma}\label{lem:orgate}
For all $y\in G_S\cap R$, $\discrep_S(\orgate(y))\leq -1$ with
equality if $S\cap L(\orgate(y))=f(S,y)$.
\end{lemma}
\begin{proof}
Consider the four check nodes $C(v)$ for some internal node $v$ of
the tree used in the construction of $\orgate(y)$. Let $u,w$ be
the children of $v$ in the original binary tree tree. Then, if
either $u,v,w\in S$ then $\discrep_S(C(v))\leq 0$ with equality
iff $v\in S$ and at least one of $u,w\in S$. Consequently, if
$\Gamma(y)\cap L' \cap S\neq \emptyset$, $\discrep_S(\cup_v
C(v))\leq 0$ with equality iff $S\cap L(\orgate(y))=f(S,y)$. In
particular, the root of the binary tree is in $S$ and therefore
the final check node $z$ has odd degree. Therefore,  if
$\Gamma(y)\cap L' \cap S\neq \emptyset$, $\discrep_S(\cup_v
C(v))\leq -1$ with equality if $S\cap L(\orgate(y))=f(S,y)$.
\end{proof}

\subsubsection{Hardness of \mingood}

Note that the graph $G$ that has been constructed has $|L|\leq
mn+2n(n-k)$ and $|R|\leq (n-m)+n^2$.

\begin{lemma}\label{lem:goodcw}
 $\discrep_S(G)\geq 0$ iff $S\cap L'$ is a codeword and for each $x\in S\cap L'$, $\zigzag(x)\cap L\subset S$.
\end{lemma}
\begin{proof}
According to Lemma \ref{lem:zigzag} and \ref{lem:orgate},
\begin{eqnarray*}
\discrep_S(G)&=&\discrep_S(G')+\sum_{x\in L'} \discrep_S(\zigzag(x))+\sum_{y\in R'} \discrep_S(\orgate(y))\\
&\leq & \discrep_S(G')-\sum_{x\in L'} I_{\zigzag(x)\cap L\not
\subset S}-|\Gamma(S)\cap R'|.
\end{eqnarray*}
Note that $\discrep_S(G')\leq |\Gamma(S)\cap R'|$ and therefore
 $\discrep_S(G)\geq 0$ implies that $d_{G_S}(y)$ is even for all $y\in R'$ and $\zigzag(x)\subset G_S$ for all $x\in S\cap L'$.
Again, according to Lemma \ref{lem:zigzag} and \ref{lem:orgate} if
$\forall y\in R',\ d_{G_S}(y)= 0 \bmod 2$, and $\forall x\in S\cap
L', \ \zigzag(x)\subset G_S$, then  $\discrep_S(G)\geq 0$.
\end{proof}

\begin{theorem}\label{thm:mingood}

For any constant $\alpha$, there is no polynomial-time,
$\alpha$-approximation algorithm for $\mingood$, unless $RP=NP$.
\end{theorem}
\begin{proof}
Assume that $S$ is a good set such that $S\leq \alpha \mingood$
for some constant $\alpha$. By Lemma~\ref{lem:goodcw} and
Lemma~\ref{lem:orgate},
\[|S|=|S\cap L'|m + \sum_{y\in \Gamma(S\cap L')} |f(S,y)|,
\]
and $S\cap L'$ corresponds to a codeword. But  $\sum_{y\in
\Gamma(S\cap L')} |f(S,y)|\leq 2n(n-k)$, and so by setting $m$
sufficiently large we get a constant approximation for \mincw. But
no such approximation exists unless $RP=NP$  \cite{DMS03}.
\end{proof}

\subsubsection{Hardness of \minctrap}
To achieve the hardness result for \minctrap we need to further
augment our graph $G$ with multiple ``\ballast{}'' constructions.
We call the resulting graph $G^+$. The intuition behind \ballast
is that no nodes from \ballast will be chosen in $S$ while the
multiple copies of \ballast will ensure that the complement of $S$
is also good. A single $\ballast$ consists of nodes $ L(\ballast)=
\{u_1, \ldots, u_{l}\}$, $R(\ballast)= \{v_1, \ldots, v_{l}, w_2,
\ldots, w_{l}\},$ and edges,
\[E(\ballast)= \{(u_i,v_{i}): i\in
[l]\}\cup \{(v_i,u_{i+1}): i\in [l-1]\}\cup \{(v_l,u_1)\}\cup
\{(u_i,w_i):1\leq i\leq l-1\}\ .
\]
We consider setting $l =n|L|$ and adding $|R|$ copies of
\ballast{} to $G$.

\begin{figure}[tb]
\centering \epsfig{file=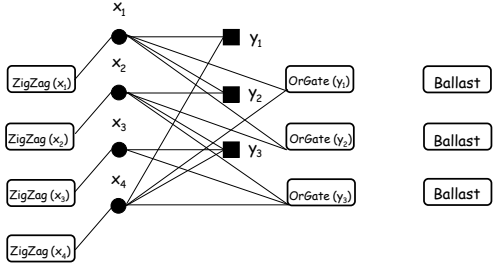, angle=-90,
width=6.5cm} \caption{Combining the $\zigzag, \orgate,$ and $\ballast$ constructions.}
\label{fig:zigzag}
\end{figure}


\begin{lemma}\label{lem:allofballast}
$\discrep_S(\ballast)\leq 1$ with equality iff $L(\ballast)\subset
S$.
\end{lemma}
\begin{proof}
Let $A=S \cap \{u_1, \ldots, u_l\}$. Note that $\Gamma(A)$
contains at least $|A|-1$ nodes with odd degree with equality iff
$L(\ballast)\subset S$. $\Gamma(A)$ contains at most $|A|$ nodes
of even degree with equality iff $L(\ballast)\subset S$.
\end{proof}

\begin{lemma}\label{lem:ubongood}
Assuming there exists a non-zero codeword, there is a good set in
$G$. Furthermore, any good set in $G$ is a trapping set for $G^+$.
\end{lemma}
\begin{proof}
Let $S'$ be the subset of $L'$ corresponding to the minimum weight
codeword. Let
\[S=S'\cup \left( \bigcup_{x\in S'} L(\zigzag(x))\right )
\cup \left (\bigcup_{y\in \Gamma(S')} f(S,y)\right )\enspace .\]
Then $S$ is a good set in $G$. For the second part of the lemma
note that by Lemma~\ref{lem:allofballast}, for $S\subset L$,
$\discrep_{\bar{S}}(G^+)\geq |R|-|R|=0$.
\end{proof}

\begin{theorem}
For any constant $\alpha$, there is no polynomial-time,
$\alpha$-approximation algorithm for $\mingood$, unless $RP=NP$.
\end{theorem}
\begin{proof}
Assume that $S$ is a trapping set such that $S\leq \alpha
\minctrap$ for some constant $\alpha$. By Lemma
\ref{lem:ubongood}, we know that $|S|\leq \alpha |L|$ and hence
$S$ does not include all left hand side nodes of any copy of
\ballast because doing so would imply that $|S|\geq
|L(\ballast)|=n |L|$. But then by Lemma \ref{lem:allofballast}, we
may assume that no nodes from \ballast are included in $S$ because
removing all such nodes from $S$ increases $\discrep_S(G)$.
Consequently $S$ must be a subset of $L$. Since any good subset of
$L$ is a trapping set, $\mingood(G)=\minctrap(G^+)$. But, by
Theorem \ref{thm:mingood}, there is no constant approximation of
\mingood.
\end{proof}

\section{Hardness of Approximation Results for Sparse Codes}
\label{sec:ldpc}

The fact that a problem is NP-hard usually does not imply that a
special instance of the problems is NP-hard. Since iterative
decoding algorithms have both linear-time complexity and offer
good decoding performance only for special classes of codes, it is
important to establish the analogues of the results in
Section~\ref{sec:main} for such codes. We provide next a set of
results establishing the hardness of approximating stopping and
trapping sets for low-density parity-check (LDPC) codes.

LDPC codes are linear block codes for which the parity-check
matrix $H$ is sparse -i.e., for which $H$ has a ``small'' number
of non-zero entries. More formally, we define an LDPC code as
follows. An LDPC code is a code with the property that each
variable and check node in its Tanner graph $G=(L \cup R, E)$ has
degree at most $\delta_v$ and $\delta_c$, respectively, for some
constants $\delta_v,\delta_c>2$ independent on $n$.


\begin{theorem}
There exists a constant $\alpha>1$ such that it is NP-hard to
$\alpha$-approximate \minstop in the Tanner graph of an LDPC code.
\end{theorem}
The proof follows along the same lines as the proof
of NP-hardness using reduction from the problem \minvc
problem~\cite{KC06}: Let $G=(V,E)$ be an undirected graph, which,
without loss of generality, can be assumed to be connected and of
vertex degree bounded from above by three. Furthermore, also
assume that $|V|=n$, $|E|=m$, and that $E=\{{e_1,\ldots,e_m\}}$,
$V=\{{v_1,\ldots,v_n\}}$. Without loss of generality, one can set
$e_1=(v_1,v_2) \in E$. A bipartite graph $G_{vc}$ is constructed
as follows: the left hand side vertices of the graph consist of
nodes $L=L_0 \cup L_1$, where $L_0=V$, and
$L_1=\{{e'_1,\ldots,e'_m\}}$. The right hand side vertices of the
graph consist of nodes $R=R_0 \cup R_1$, with $R_0=E$, and
$R_1=\{{z_1,\ldots,z_m\}}$. The set of edges of $G_{vc}$ is a
collection of ordered pairs the following form:
\begin{align*}
&~\{(e_i \in R_0,u \in L_0), (e_i \in R_0,v \in L_0): e_i=(u,v)
\in E\} \cup \{(e_i \in R_0,e'_i \in L_2): 1 \leq i \leq m \} \cup
\\
&~\{(z_i \in R_1, e'_i \in L_1), (z_i \in R_1, e'_{i+1} \in L_1):
1 \leq i \leq m-1 \} \cup \{(z_m \in R_1,v_1 \in L_0),(z_m \in
R_1,e'_1 \in L_1)\}.
\end{align*}
It is straightforward to show that if $S$ is a stopping set in
$\mathcal{G}$, then $S \cap L_0$ is a vertex cover in
$\mathcal{G}$~\cite{KC06}. As a consequence, there exists a
constant $\epsilon>0$ such that there is no $(1+\epsilon)$
approximation algorithm for the \minstop problem, unless P=NP.

Note that in the construction, each vertex in $L$ has degree
bounded from above by four (the auxiliary variable node
$e_1',\ldots,e_{|E|}$ have, by construction, degree two, while all
vertices in $V$ other than $v_1$ and $v_2$ have degree at most
three; the vertices $v_1$ and $v_2$ can have degree at most four).
Similarly, the check nodes have maximum degree three, since by
construction, the vertices $z_1,\ldots,z_{|E|}$ have degree two,
while the vertices in $R_0$ have degree three.

One can establish the even stronger result that the \minstop
problem for LDPC codes remains NP hard even for codes with Tanner
graphs that avoid cycles of length four. This follows from the
same arguments used in the proof of the theorem above, with an
additional reference to the hardness of the \minscinter problem,
which also holds in the setting of sparse codes~\cite{kumar}.


\begin{theorem}
There exists a constant $\alpha>1 $ such that it is NP-hard to
$\alpha$-approximate \minetrap in the Tanner graph of an LDPC
code.
\end{theorem}
\begin{proof} The proof follows along the same lines as the proof
of Theorem~\ref{th:mine}, with the three-dimensional matching
problem replaced by its constraint version involving a bounded
number $\ell$ of appearances of each element in $X$.
\end{proof}


\begin{theorem} The problems \mld and \mincw are
NP-hard for LDPC codes.
\end{theorem}
\begin{proof} The proof is a direct consequence of the fact that the
parity-check matrix used in the reduction from the \mtdm to the
\mld problem is sparse (it has column weight three, and the row
weight can be made bounded as well by invoking the constraint that
any element of $X$ cannot appear more than $r \geq 3$ times). The
claimed result follows from the observation that there exists a
polynomial-time reduction algorithm from the \mld to the \mincw
problem~\cite{vardy-1-97,vardy-2-97}.
\end{proof}

As a consequence of the above finding, all trapping set problems
described in Section~\ref{sec:main}, for which the hardness was
established in terms of reductions from the \mincw problem, remain
NP-hard for the class of LDPC codes.
%

\section{Estimation of the Error-Floor}
\label{sec:errorfloor}

The error floor is a phenomena inherent to iterative decoders that
manifests itself as a sudden change in the slope of the BER
performance of a code. Alternatively, it represents a phase
transition in the dynamical system of the decoder that prohibits
it from attaining a sufficiently low BER. The error floor usually
appears at moderate to high signal-to-noise ratios, i.e. for small
values of the erasure and error probability $p$ of the BEC and BSC
channel. For such values of $p$, the codeword error-rate $R(p)$
has the form
\begin{equation}
\log\left(R(p)\right) \simeq \log(N_{\kappa})+\kappa\;\log(p),
\end{equation}
where $\kappa$ denotes the size of the smallest stopping/trapping
sets, while $N_i$ represents the number of such sets. The
dominating term in the expression is the linear term
$\kappa\;\log(p)$.

As a consequence of the results in Section~\ref{sec:main}, we have
the following result.

\begin{corollary} Unless $P=NP$, there is no polynomial time algorithm
for estimating the error-floor of codes used over the BEC and BSC
within an $O(1)$ term.
\end{corollary}

For the AWGN channel with noise variance $\sigma^2$, a heuristic
formula for the codeword error-rate was derived in~\cite{rich03},
where it was shown that
\begin{equation}
R(\sigma) \geq \sum_{T \in \mathcal{T}}P(T,\sigma), \notag
\end{equation}
where $\mathcal{T}$ denotes the set of dominant (small) elementary
trapping sets for the given code, and $P(T,\sigma)$ is the
probability of decoder failure on a trapping set $T$. It was
observed that simulation of decoding can be viewed as stochastic
process for finding trapping sets~\cite{rich03}. This, and other
methods that rely on combining simulation techniques with ``aided
flipping'' methods and greedy search strategies, were all observed
to be inefficient when estimating the error-floor of ``good
codes'' - i.e. codes with large minimum stopping and trapping set
sizes. In the next section, we show that some problems discussed
in the paper has complexity that grows exponentially with the size
of the smallest set being sought, but only polynomially with
respect to the size of the input (i.e., code length).
Consequently, one can easily find the smallest stopping sets of
fairly long codes, provided that the size of such stopping sets is
not greater than $10-15$~\cite{downey97,downey99,rosnes07}. This
was observed in several papers, including~\cite{rosnes07}.

\section{Fixed-Parameter Tractability}
\label{sec:fpt}
Parameterized complexity represents a measure of the computational
cost of problems that have several input parameters. Problems for
which one of the parameters, say $\pi$, is fixed are called
parameterized problems. There exist problems that require
exponential running time in the parameter $\pi$ but that are
computable in a time that is polynomial in the input size. Hence,
if $\pi$ is fixed at a small value, such problems can still be
\emph{exactly} solved in an efficient manner. A parameterized
problem that allows for the existence of such polynomial time
algorithms is termed \emph{a fixed-parameter tractable} problem
and it belongs to the class FPT, first studied by Downey and
Fellows~\cite{downey99}.

Many NP-complete problems are fixed-parameter tractable. As an
example, the \minvc is FPT, with complexity $O(\kappa \,
n+(4/3)^{\kappa}\,\kappa^2)$, where $\kappa$ denotes the size of
the smallest vertex cover, and $n$ is the size of the input, i.e.,
the number of vertices in the graph. Despite the fact that \minvc
is a special instant of \minhs with set sizes equal to two, the
latter is not known to have FPT algorithms when parameterization is
performed only with respect to the size of the smallest hitting
set $\kappa$. Strong evidence suggests that such an algorithm does
not exist, since \minhs is $W[2]$-complete (for the non-trivial
definition of the $W[2]$ class, see~\cite{downey97}). It is only
known that \minhs is FPT when the set sizes are bounded, and
parameterization is performed with respect to, say,
$\kappa+\delta_{\text{max}}$, where $\delta_{\text{max}}$ denotes
the size of the largest set in the \minhs formulation.

In this section, we use the results
of~\cite{cesati06,fernau05,rosamond05} to show that the \mincstop
problem is FPT. Furthermore, by invoking the recent results
in~\cite{damaschke07}, we show that the problem of enumerating
\emph{all} cover stopping sets is FPT as well.

\begin{theorem} The problem \mincstop for LDPC codes of maximal constant check node
degree $\delta_c$ is in FTP, with best known complexity bound of
the form
\begin{equation}
O\left( \left(
\frac{\delta_c-2}{2}\left(1+\sqrt{1+\frac{4}{(\delta_c-2)^2}}\right)\right)^{\kappa}+n\right).
\end{equation} \label{fpt-1}
\end{theorem}
The algorithm that achieves this bound is a tree search algorithm,
see~\cite{fernau05}.
\begin{theorem} The problem of enumerating all minimal cover
stopping sets in LDPC codes of maximal constant check node degree
$\delta_c$ is in FTP, with best known complexity bound of the form
$
O^{\star}\left( (\delta_c-1+o(1))^{\kappa}\right),
$ \label{fpt-2}
where $O^{\star}$ refers to an $O(\cdot)$ function for which all
polynomial factors are suppressed, and where $\kappa$ stands for
the size of the smallest cover stopping set.
\end{theorem}
As a final remark, the problem \minstop can be shown to be
W[1]-hard, due to its connection to the Exact Even Set
problem~\cite{cesati06}.

\section{Conclusion} \label{sec:conc}

We showed that a class of problems, pertaining to the size of the
smallest stopping and trapping sets in Tanner graphs is NP-hard to
even approximate. Furthermore, we showed that similar results
apply to the class of LDPC codes. Our findings provide one of the
few known \emph{families} of codes for which the minimum distance
and stopping set problems are NP-hard. We also show that a simple
instance of the stopping set problem for LDPC codes, namely the
\emph{complete} stopping set problem, is fixed parameter
tractable.

\end{document}